\documentclass[9pt, conference]{IEEEtran}
\IEEEoverridecommandlockouts
\usepackage{cite}
\usepackage{amsmath,amssymb,amsfonts}
\usepackage{algorithmic}
\usepackage{graphicx}
\usepackage{textcomp}
\usepackage{xcolor}
\usepackage{microtype}
\usepackage{subfigure}
\usepackage{algorithm}
\usepackage[font=small]{caption}
\usepackage{amsmath}
\usepackage{breqn}

\newcommand\blfootnote[1]{%
  \begingroup
  \renewcommand\thefootnote{}\footnote{#1}%
  \addtocounter{footnote}{-1}%
  \endgroup
}

\def\BibTeX{{\rm B\kern-.05em{\sc i\kern-.025em b}\kern-.08em
    T\kern-.1667em\lower.7ex\hbox{E}\kern-.125emX}}
\begin{document}

\title{Generation \& Evaluation of Adversarial Examples for Malware Obfuscation
}

\author{
\IEEEauthorblockN{Daniel Park}
\IEEEauthorblockA{\textit{Rensselaer Polytechnic Institute} \\
Troy, NY \\
parkd5@rpi.edu}
\and
\IEEEauthorblockN{Haidar Khan}
\IEEEauthorblockA{\textit{Amazon Alexa} \\
New York, NY \\
khhaida@amazon.com}
\and
\IEEEauthorblockN{B{\"u}lent Yener}
\IEEEauthorblockA{\textit{Rensselaer Polytechnic Institute} \\
Troy, NY \\
yener@cs.rpi.edu}
}

\maketitle

\begin{abstract}
There has been an increased interest in the application of convolutional neural networks for image based malware classification, but the susceptibility of neural networks to adversarial examples allows malicious actors to evade classifiers. Adversarial examples are usually generated by adding small perturbations to the input that are unrecognizable to humans, but the same approach is not effective with malware. In general, these perturbations cause changes in the byte sequences that change the initial functionality or result in un-executable binaries. We present a generative model for executable adversarial malware examples using obfuscation that achieves a high misclassification rate, up to 100\% and 98\% in white-box and black-box settings respectively, and demonstrates transferability. We further evaluate the effectiveness of the proposed method by reporting insignificant change in the evasion rate of our adversarial examples against popular defense strategies.
\end{abstract}

%\IEEEpeerreviewmaketitle
\blfootnote{\fontsize{7.5}{8}\selectfont This research was done while H. Khan was a graduate student at Rensselaer Polytechnic Institute. D. Park is supported by the US Department of Defense's Science, Mathematics and Research for Transformation (SMART) Scholarship for Service Program.}

\section{Introduction} \label{intro}
With the growth of the Internet, the number of malware has been rapidly increasing. Machine learning has significantly improved existing solutions in malware detection, classification, and network intrusion detection and prevention.  Static and dynamic analysis have been the cornerstone of malware detection and classification, but are increasingly being replaced by rule based and machine learning models trained on features extracted from analysis, such as $n$-grams \cite{Santos09}. This led to work, such as that of \cite{Kinable10} and \cite{OKane16}, in extracting richer sets of features and representations directly from malware.

It has been proposed to use the image representation of malware as input, allowing use of image processing and classification techniques for malware classification \cite{Nataraj11}. Features extracted from malware images were popularly used during a 2015 competition~\cite{microsoft-kaggle}, resulting in significant improvement for the top performing classifier~\cite{notooverfitting15}. The methods proposed by \cite{Zhou17} and \cite{Gibert2019} only used malware image features and achieved high classification accuracy.

However, \cite{Szegedy13} found that neural networks are vulnerable to adversarial examples. Adversarial examples are crafted by perturbing valid inputs such that they are indistinguishable from the valid inputs when viewed by humans but are confidently misclassified by a machine learning classifier. As the concept of adversarial examples has mostly been explored in the domain of natural images, interest has shifted to further research and understanding of adversarial examples in the malware domain \cite{Kreuk19, Grosse2017}. This immediately led us to question the effectiveness of adversarial examples for image representations of malware. 

In this paper, we are interested in the application of machine learning to malware classification and the effect that adversarial examples have in this regard. Similar to adversarial examples in the natural image domain, an adversarial example of malware is an \textbf{executable} binary file that is crafted from an existing malware sample with the purpose of fooling a machine learning based classifier. However, adversarial malware examples must also preserve the original functionality while being executable. This paper presents a novel approach to creating executable adversarial examples by minimally modifying malicious raw binaries using a dynamic programming based insertion algorithm that obfuscates the \textit{.text} section of a binary with executable byte sequences. 

We report the misclassification rate of our adversarial malware examples against popular machine learning classifiers. We also test the effectiveness of the examples by strengthening the classifiers using popular defensive techniques, adversarial training and distillation, while comparing our proposed method with the current literature. We also show that dummy code insertion patterns can be learned and used as heuristics for obfuscation without use of machine learning based approaches.

\subsection{Contributions}
The main contributions of this paper are as follows:
\begin{itemize}
    \item We present a novel approach for generating adversarial malware executables and obfuscation using malware images.
    \item We show transferability of our generated adversarial examples to other classifiers.
    \item We evaluate the proposed method and related work against defended neural networks using adversarial training and distillation.
    \item We compare obfuscation generated using the proposed method to other common obfuscation techniques.
\end{itemize}
% TODO

\subsection{Organization}
This paper is organized as follows. Section \ref{background} gives a brief background on malware classification and adversarial examples. Section \ref{Methodology} describes the mechanisms used in the proposed method for generating executable adversarial malware examples. Section \ref{Experimental results} reports results under white-box and black-box settings. It also reports the results after strengthening neural networks using adversarial training and distillation. In Section \ref{Discussion}, we discuss interesting aspects of the work, as well as possible extensions of the proposed method. Section \ref{Related works} presents related work in adversarial example generation for malware and obfuscation. We conclude the paper in Section \ref{conclusion}.

\section{Background}\label{background}
In this section, we describe the relevant background and malware classification setup. We also discuss the nature of adversarial examples in the malware image domain.

    \subsection{Convolutional neural networks}
    Convolutional neural networks (CNN) are often the model of choice when the input data contains spatial or temporal relationships, such as pixels in an image or samples in a time series. The learnable filters in a CNN are designed to explicitly capture the spatial or temporal invariances in the data. This, combined with the layered representation learning of deep neural networks allows CNN to effectively model many different types of data.
    
    A typical CNN consists of multiple convolutional layers with pooling layers interspersed in between. The intuition behind this architecture is that the layers learn specialized representations of the input data for the task of interest. 
    
    CNN's use in natural image classification has led to their use in malware classification. In the following section, we discuss an image representation of malware that can be used as input to a CNN.
    
    \subsection{Data representation}
    In this work, we represent Windows malware binaries as grayscale images generated from the bytecode. 
    
    A whole malware binary is read into a 2-dimensional array as 8-bit unsigned integers. Because each of these 8-bit unsigned integers are between 0 and 255, they can be visualized as a grey-scale image where 0 is black and 255 is white.

    Previous work has shown that this image representation of malware can be used to see shared repeated patterns and distinct characteristics between binaries in the same class \cite{Nataraj11}.
    
    In this work, we used the Malimg dataset introduced in \cite{Nataraj11}. In addition to the Malimg dataset, this work also uses the Microsoft malware challenge dataset \cite{microsoft-kaggle}. For this dataset, the width of the image was fixed according to the recommended width for the largest malware sample. 

    \subsection{Adversarial examples \& malware}
    The popularity of the image representation of malware has led to increased usage of deep learning techniques for classifying malware at high accuracies \cite{Yue17, Kabanga18}.
    
    Adversarial examples of malware images, generated using Fast Gradient Signed Method (FGSM) or the Carlini and Wagner (C\&W) method can be used to evade classification \cite{Goodfellow14, Carlini16}. However, adversarial malware examples must also be executable and have the same effective program logic. A perturbation on a malware image does not necessarily ensure these properties. For example, let $x = [137, 200]$, two pixels representing a binary sequence. When converted to hexadecimal and then to x86 instructions, we get $89c8$ and \textit{mov eax, ecx}, respectively. Consider a small perturbation on the image which results in $x' = [137, 201]$. When $x'$ is converted into assembly instructions, we get \textit{mov ecx, ecx}, which does not produce the same results as $x$.
    
    \subsubsection{Generating adversarial examples and noise}
    
    Adversarial examples are generally natural images, where a perturbation of the original image causes it to be misclassified even though it appears to be a copy of the original. We define adversarial examples and the methods used to generate them below before introducing the proposed method for generating executable adversarial examples.

    Given an input binary file $x$, an adversarial example generated using $x$ is a perturbed version of $x$, $x'$, such that 
    \begin{equation}
        x' = x + \delta  
    \end{equation}
    where $\delta$ is some additive perturbation, which causes a classifier to incorrectly label the binary file while preserving the functionality of the original file. $x'$ can be generated using various attack techniques. In this work, we use the FGSM \cite{Goodfellow14} and C\&W attacks \cite{Carlini16}.
    
    In a non-targeted attack as explained in Section \ref{threat}, the goal is to cause misclassification. The task for the adversary is to generate an $l_2$-bounded perturbation of a binary $x$ with true class $y$ such that 
    \begin{equation}
    \label{eqn:x-x}
        ||x' - x||_2 < \epsilon
    \end{equation}
    where $x'$ is the modified binary, $||\cdot||_2$ is the $l_2$-norm, and $x'$ is executable and preserves the functionality of $x$ while having a $l_2$ distance of less than a hyper-parameter $\epsilon$.
    
    In a targeted attack, Equation \ref{eqn:x-x} still applies. The only additional requirement is that the adversarial example is misclassified as a target class $y'$ chosen by the adversary.
    
    \subsection{Defensive techniques}
    In this section, we provide a brief summary of the defensive techniques against adversarial examples that will be used later to assess our method's effectiveness.
    
        \subsubsection{Adversarial training}
        Reference \cite{Szegedy2014} showed that training on a combination of discovered adversarial examples and clean examples somewhat regularizes the neural network. The goal of adversarial training is to learn to correctly classify future inputs that have undergone adversarial perturbations. Reference \cite{Goodfellow14} showed that adversarial training on the MNIST dataset reduced the original adversarial examples' misclassification rates from $89.4\%$ to $17.9\%$. Adversarial examples generated using the adversarially trained model had an evasion rate of $40.9\%$.
        
        \subsubsection{Distillation}
        Distillation was originally introduced by \cite{Hinton15} as a method to reduce the size of deep neural network architectures while preserving knowledge acquired during training and extended by \cite{Papernot15} to defend against adversarial examples. Intuitively, neural network distillation is the process of extracting class probability vectors from a first deep neural network to train a second deep neural networks of reduced dimensionality. This is based on the notion that knowledge is encoded in a model's output probability vector in addition to its weights.

    \subsection{Obfuscation}
    The main goal of obfuscation is to make a program difficult to understand while preserving the logic of the original program. In this preliminary work, we focus on generating adversarial examples using an obfuscation technique called dummy code insertion.
    
\begin{figure}[h!]
    \centering
    \begin{subfigure}
        \centering
        \includegraphics[width=.35\linewidth]{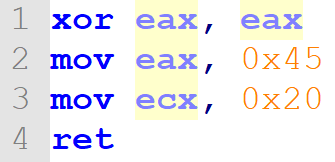}
    \end{subfigure}
    \begin{subfigure}
        \centering
        \includegraphics[width=.35\linewidth]{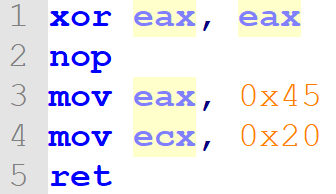}
    \end{subfigure}
    \caption{On the left, we have a simple assembly program that zeros out the $eax$ register before 
    moving the hex numbers 45 and 20 into registers $eax$ and $ecx$, respectively. On the right, we have 
    another program that has the same functionality, but has a $nop$ or no-operation code inserted.}
    \label{fig:dummy}
    \end{figure}

    Dummy code insertion introduces code sequences that do not affect the program logic. These sequences are named after the \textit{nop} or \textit{no operation}. There are multiple ways to insert \textit{nops} into a program, in addition to simply inserting the \textit{nop} instruction as seen in Figure \ref{fig:dummy}. These are called semantic \textit{nops} and are instructions or sequences of instructions that do not affect program logic. 

%%%%%%%%%%%%%%%%%%%%%%%%%%%%%%%%%%%%%%%%%%%%
%% Dummy Code Example
%%%%%%%%%%%%%%%%%%%%%%%%%%%%%%%%%%%%%%%%%%%%   
    
%%%%%%%%%%%%%%%%%%%%%%%%%%%%%%%%%%%%%%%%%%%%
%% Should be at the beginning of methods
%%%%%%%%%%%%%%%%%%%%%%%%%%%%%%%%%%%%%%%%%%%%
\begin{figure*}[ht!]
    \centering
    \includegraphics[width=.72\linewidth]{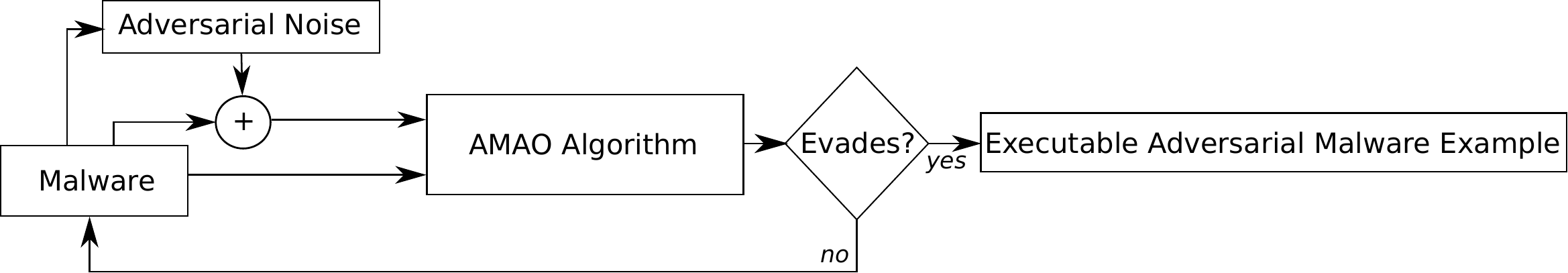}
    \caption{This block diagram outlines the full proposed method for creating executable adversarial malware examples. The initial input is a malware sample. Using the image representation of the input malware, adversarial noise is generated using FGSM or the C\&W attack for a targeted or untargeted attack. The adversarial noise is added to the malware image, which may or may not be executable or have the same functionality. The output of this addition will be referenced to as the image adversarial example. The image adversarial example and the input malware are inputs to the proposed AMAO obfuscation algorithm to generate an executable adversarial malware example. If the output of the algorithm does not evade a target classifier, we repeat the procedure.}
    \label{fig:block}
\end{figure*}
%%%%%%%%%%%%%%%%%%%%%%%%%%%%%%%%%%%%%%%%%%%%
%%%%%%%%%%%%%%%%%%%%%%%%%%%%%%%%%%%%%%%%%%%%
%%%%%%%%%%%%%%%%%%%%%%%%%%%%%%%%%%%%%%%%%%%%

\section{Methodology}\label{Methodology}
In this section, we will formally define our problem and threat model. We also describe how the Fast Gradient Sign Method and the Carlini \& Wagner method can be applied to generate \textbf{non-executable} adversarial examples from malware images. We introduce a dynamic programming based algorithm, Adversarial Malware Alignment Obfuscation (AMAO), to achieve this goal. A general block diagram of the full method is given in Figure \ref{fig:block}.

    \subsection{Threat model}\label{threat}
    In this section, we define the adversary's goal, capabilities, and attack surface. 
    
    \subsubsection{Adversarial Goal}
    In this work, the goal of the adversary is to generate malware samples that evade classification by deep neural network classification models. In an untargeted attack, the adversary adjusts the malware sample to be classified as any class but the true class. In a targeted attack, the adversary adjusts the malware sample to be classified as a chosen false class.
    
    \subsubsection{Adversarial Capabilities}
    The adversary does not have any training phase capabilities, e.g, data injection and data modification, and limits the attack surface to the testing phase. We consider two separate types of adversaries, a white-box and a strict black-box adversary. The white-box adversary has total knowledge about the target model's architecture, weights, training algorithm, and training data distribution. The strict black-box adversary only has the ability to collect input-output pairs $(\mbox{malware}, \mbox{class}')$ where $\mbox{class}'$ is the label assigned the the sample by the target classifier.

%%%%%%%%%%%%%%%%%%%%%%%%%%%%%%%%%%%%%%%%%%%%
%% Image should be after block diagram
%%%%%%%%%%%%%%%%%%%%%%%%%%%%%%%%%%%%%%%%%%%%
    \begin{figure*}[h!]
        \centering
        \hspace{4mm}\includegraphics[width=.60\linewidth]{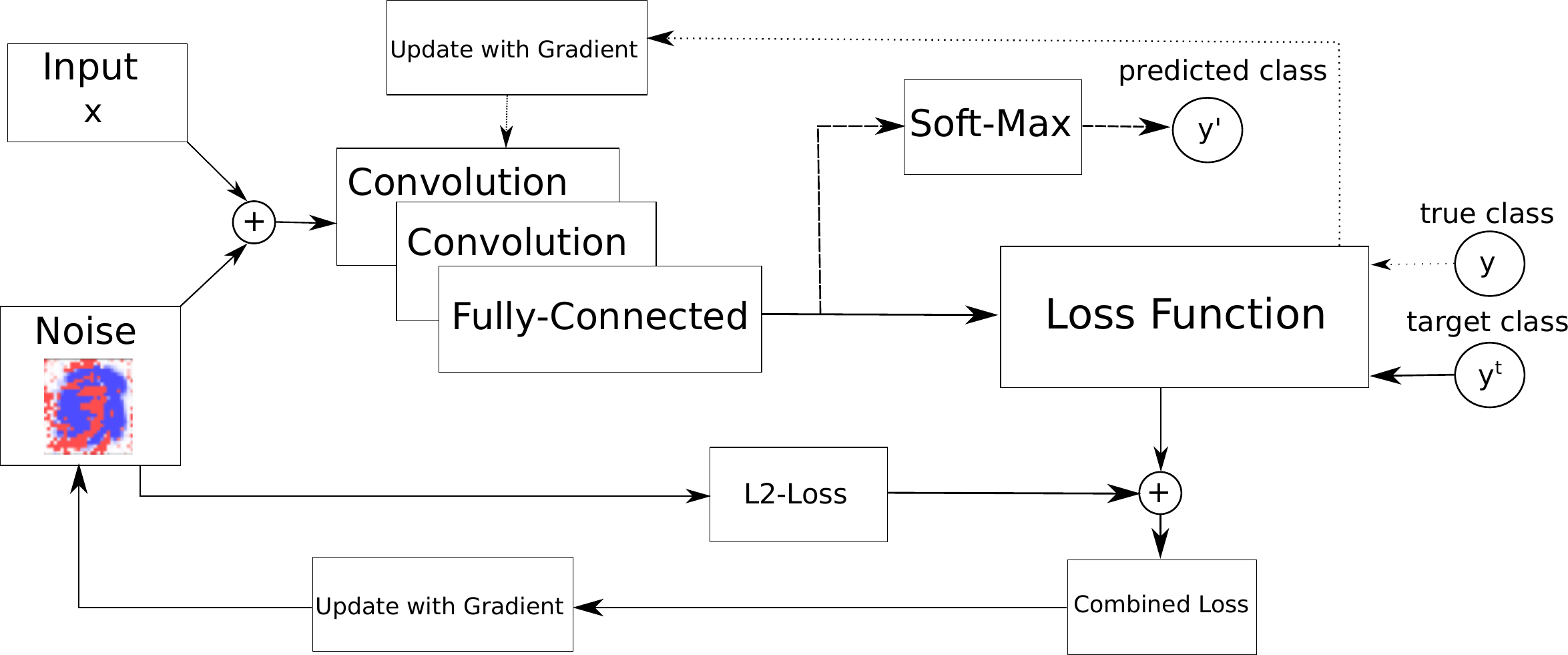}
        \caption{This is the architecture of a fully trained convolutional neural network that is used to generate adversarial noise. During the noise optimization process, we begin with Gaussian noise added to valid inputs before adjusting that noise such that the now perturbed input is misclassified (as a target class $y'$ if it was provided). The optimizer for the adversarial noise minimizes both the normal loss-measure and the $l_2$-loss measure to cause misclassification but to also bound the perturbation.}
        \label{img:architecture}
    \end{figure*}    
%%%%%%%%%%%%%%%%%%%%%%%%%%%%%%%%%%%%%%%%%%%%
%%%%%%%%%%%%%%%%%%%%%%%%%%%%%%%%%%%%%%%%%%%%
%%%%%%%%%%%%%%%%%%%%%%%%%%%%%%%%%%%%%%%%%%%%   

    \subsection{Semantic \textit{nops}}  
    We compiled a list of 23 semantic \textit{nops} that are used in the dynamic programming algorithm introduced in Section \ref{amao} to create our executable adversarial examples. The semantic \textit{nops} are inserted into a malware sample to create executable adversarial malware examples using a dynamic programming algorithm.
    
    \subsection{Generating executable adversarial malware examples} \label{amao}
    The Adversarial Malware Alignment Obfuscation (AMAO) algorithm follows a standard dynamic programming structure and resembles string matching algorithms. The algorithm takes two binary strings, $B1$ and $B2$, and a list of insertion points as inputs. The binary string $B1$ has a length of $n$ (in bits) and is the binary representation of the non-executable adversarial example generated by a standard adversarial example generation attack (such as C\&W). $B2$ has length $m$ (in bits) and is the binary representation of the original malware sample. Insertion points are address locations within $B2$ where an insertion of a semantic \textit{nop} would not split an existing opcode and its operands. The goal of AMAO is to insert semantic \textit{nops} at $B2$'s insertion points such that $B2$ matches $B1$ as closely as possible. Note that $m < n$ as we are only interested in inserting semantic \textit{nops} in the original malware and are not modifying the non-executable adversarial example. The output of the algorithm is an adversarial malware example that follows the original sample's program logic.
    
    We define an $n \times m$ table $O$ for memoization. $O[i][j]$ is defined as the minimum distance between $B1[1:i]$ and $B2[1:j]$ with $B1[i]$ matched with $B2[j]$. We observe that the following recurrence holds: 
    
    \begin{equation}
    \label{eq:subproblem}
        O[i][j] = \min \left\{
                            \begin{array}{l}
                                 O[i-1][j-1] \\
                                 minCode
                            \end{array}
                        \right. + D(B1[i], B2[j])
    \end{equation}
    
    where
    
    \begin{multline}
        minCode = \min_{c \in L_{\mbox{nops}}}\{O[i][j-len(c) -1] + \\ D(B1[i-len(c):i], c) \}
    \end{multline}
    
    for some distance function $D$ between binary strings and a list of semantic \textit{nops} $L_{\mbox{nops}}$. At each step, the algorithm chooses between inserting a semantic \textit{nop} and not inserting anything. We complete our description of the algorithm by noting that the base cases are trivial; $O[i][1]$ for all $i=1\ldots n$ is equal to $D(B1[i], B2[1])$. After the base cases are computed, we complete the rest of the table using the sub-problem defined in Equation \ref{eq:subproblem} and Algorithm \ref{alg:dynamic}. Note that table $O$ is lower triangular as we assume $j \geq i$. 
    
    The distance metric was the sum over all bit differences, or
    \begin{equation}
        D(x_1, x_2) = \sum_{i=0}^n \left\{
                                        \begin{array}{ll}
                                            0 & \quad x_1[i] = x_2[i] \\
                                            1 & \quad x_1[i] \ \ne \ x_2[i]
                                        \end{array}
                                    \right.
    \end{equation}
    
    where $x_1$ and $x_2$ are the adversarial example and malware sample, respectively, read in as a binary string. The binary strings are assumed to have the same length. In the case that $x_1$ and $x_2$ have different lengths, we pad the front of the shorter binary string with 0's.
    
%%%%%%%%%%%%%%%%%%%%%%%%%%%%%%%%%%%%%%%%%%%%
%% Edit algorithm to just point at eq 9, 10
%%%%%%%%%%%%%%%%%%%%%%%%%%%%%%%%%%%%%%%%%%%%
    \begin{algorithm}[h]
        \caption{Adversarial Malware Alignment Obfuscation \small (AMAO)}
        \label{alg:dynamic}
        \begin{algorithmic}
            \STATE{\bfseries Input: } adversarial example $B1$, original malware $B2$, insertion points $L_{ins}$, \textit{nops} list $L_{\mbox{nops}}$, distance function $D(\cdot,\cdot)$ 
            \FOR{$i\in$ range$(1, len(B1))$}
                \FOR{$j\in$ range$(i, len(B2))$}
                    \STATE Calculate $O[i][j]$ using Equation \ref{eq:subproblem}
                \ENDFOR
            \ENDFOR
        \end{algorithmic}
    \end{algorithm}  
   
   At the completion of the algorithm, the optimal solution's distance from the adversarial example is located in $O[n][m]$. From this, we can trace-back to the binary string that the optimal solution represents. 
   
   It is important to note that adding semantic \textit{nops} is much easier given the source code as the obfuscation can be added in the form of LLVM bitcode \cite{Junod15}. Without the source, we rely on the analysis and patching techniques laid out in \cite{Smithson10} and \cite{Meng16}.

    \subsection{Closed loop model}
    Because our method obfuscates a malware sample to match an adversarial example as much as possible, it is possible that the output executable may not fool the classifier. Thus, we propose a closed loop model, in which we continue to train the adversarial noise until a 0\% classification accuracy is achieved.
    
    Our initial adversarial examples are generated using the architecture shown previously in Figure \ref{img:architecture} with FGSM. Initially, the adversarial noise is generalized over the test set and may not create adversarial examples when added to a malware sample. In subsequent loops, we begin to apply more computationally expensive adversarial example generation algorithms by generating adversarial noise specific to each malware sample and creating adversarial examples using the C\&W attack. The initial generalized approach reduces computational cost, however, any adversarial example generation algorithm can be used before input to the proposed AMAO algorithm. 

    \subsection{Binary string distance metric}
    There are multiple choices for the closeness measure used between the image adversarial example and the executable adversarial malware example. We experimented with two measures, $L_2$ and $L_0$ distance. In measuring the distance between two byte-sequences, the $L_2$ distance is the standard Euclidean distance between the image representations of the two byte-sequences. The $L_0$ distance counts how many bytes differ between the two sequences. For example, let $B_1$ = 0x9000 and let $B_2$ = 0x9100. The first bytes of $B_1$ and $B_2$ differ, resulting in a distance of 1. Comparatively, the $L_2$ distance is calculated using pixel values and the distance between bytes 0x90 and 0x91 is shorter than between bytes 0x90 and 0xFF. A distance metric using the $L_\infty$ norm was not considered due to its behavior in insertion problems. This is because minimizing the maximum change to a byte between two byte-sequences does not guarantee minimal insertions between the two sequences. For example, let $B_1$ from the previous example be our goal sequence. This means that we will be inserting semantic \textit{nops} to $B_2$ such that the distance between $B_1$ and $B_2$ is minimized. Using the $L_\infty$ distance, consider the following cases. 1) $B_2$ = 0x9090; In this case the $L_\infty$ distance is 90. 2) $B_2$ = 0x90908585; the $L_\infty$ distance is also 90. If 0x85 was a semantic \textit{nop}, this byte can be inserted to a sequence an infinite number of times without affecting the distance. 

    Using the $L_2$ distance resulted in executable malware examples with a higher misclassification rate compared to those created using the $L_0$ distance. The $L_2$ distance measure allows AMAO to minimize perturbations to pixel values, which is the input to our target classifiers. The $L_0$ distance does not capture important differences between two byte-sequences, i.e., given two semantic \textit{nops} of the same length, the $L_0$ distance does not capture the difference between inserting one semantic \textit{nop} or the other. 

\section{Experimental results}\label{Experimental results}
In this section, we will outline and discuss the experimental results of our method on different classifiers in a white-box and black-box setting.

    \subsection{Dataset}
    We evaluated the effectiveness of our method using the Malimg dataset \cite{Nataraj11} and the MMBIG dataset \cite{microsoft-kaggle}. The Malimg dataset is already in the necessary black and white image representation and contains 25 unique classes. We transformed the MMBIG dataset into the correct format by parsing the given hex dumps. The MMBIG dataset consists of 9 unique classes. 
    
    To compare against random obfuscation, We used random \textit{nop} insertion as there is no algorithmic approach to dummy code insertion, to the best of our knowledge.
    
    \subsection{White-box setting}
    In this section, we consider the results of our method with access to the classifier's computational graph and weights.
    \begin{figure}[h!]
        \centering
        \includegraphics[width=.99\linewidth]{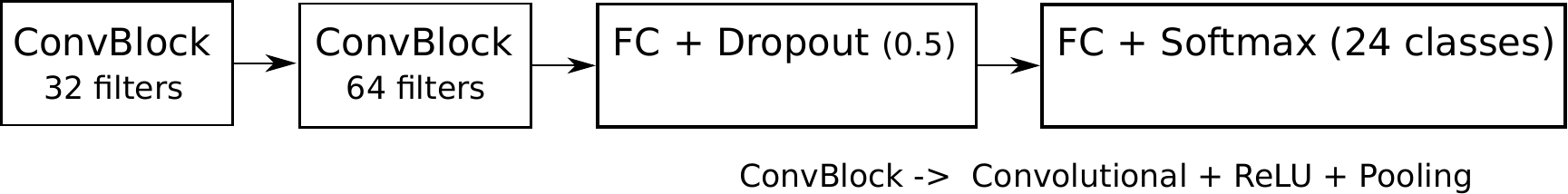}
        \caption{This is the simple CNN architecture used in the white-box experiments. The convolutional layers use a kernel size of 5, pool size and stride of 2. The fully connected layer has 128 units and the dropout layer use a rate of 0.5. The adversarial examples generated using this CNN are also used to fool classifiers in a black box setting to demonstrate transferability.}
        \label{fig:simpleCNN}
    \end{figure}
    
    We trained a simple convolutional neural network, depicted in Figure \ref{fig:simpleCNN} to classify the Malimg dataset with a classification accuracy of $90\%$ on the training validation set. This translated to an $85\%$ accuracy rate on the test set. We optimized adversarial noise for each of the 25 classes using the gradients of the trained model in a closed loop, which consists of FGSM and C\&W. In addition to the simple convolutional neural network, we also trained LeNet5 \cite{Lenet} because it was originally designed to be used with black and white characters such as MNIST. LeNet5 was trained and achieved $92\%$ classification accuracy on the training validation set and $87\%$ classification accuracy on the test set. The same procedure was used to generate white-box executable adversarial malware examples targeting LeNet5.  
    
    Our method was designed to terminate at a $0\%$ accuracy rate, or after some number $k$ iterations. As expected, our generated adversarial examples were all misclassified by the previously trained classifiers.

    \subsection{Black-box setting}
    There will always be cases when adversarial examples must be crafted without uninhibited access to the model and its weights. To evaluate or method in these circumstances, we first trained two models on the Malimg dataset. \textbf{1.)} The first is InceptionV3, which is Google's convolutional neural network architecture for image recognition introduced in \cite{Szegedy16}. InceptionV3 was trained using transfer learning to classify the Malimg dataset. \textbf{2.)} The second is a gradient boosted tree classifier, using XGBoost, based on top MMBIG Kaggle competition submissions, using popular op-codes, the top 500 4-grams, and features extracted from the image representations of the malware \cite{xgboost}. We then fed the same adversarial examples generated during our white-box setting attack, to these two classifiers without any further optimization or alteration. We also repeat this process against MalConv, a deep neural network classifier that takes raw byte sequences as input \cite{Raff17}. 
    
    \begin{table}[ht]
    \caption{Classification accuracies during the white-box and black-box setting experiments when using different obfuscation methods before input. We report accuracies using no obfuscation, one loop (AMAO$_{1}$) of the proposed method, the full proposed method (AMAO$_{\mbox{f}}$), and randomly inserted semantic \textit{nops}. For the randomized insertions, we record the range over 10 experiments. The top half of the table report results using white-box attacks and the lower half report results using black-box attacks.}
    \label{table:results}
    \centering
    \begin{tabular}{lcccr}
    \hline
     Model  & None & AMAO$_{1}$ & AMAO$_{ \mbox{f}}$ & Random \\ \hline \hline
     Simple CNN  & 85.1\% & 44.4\% & 0.0\% & 20-30\% \\ 
     LeNet5 & 86.55\% & 46.0\% & 0.0\% & 20-30\% \\ \hline
     Boosted Trees  & 99.0\% & 1.98\% & 0.0\% & 80-85\% \\ 
     InceptionV3  & 86.1\% & 50.0\% & 26.95\% & 40-50\% \\ 
     MalConv    & 99.0\% & 57.89\% & 0.0\% & 80-90\% \\
    \end{tabular}
    \end{table}
    
    %%%%%%%%%%%%%%%%%%%%%%%%%%%%%%%%%%%%%%%%%%%%%%
    %% Table used in subsection against defenses 
    %%%%%%%%%%%%%%%%%%%%%%%%%%%%%%%%%%%%%%%%%%%%%%
    \begin{table}[h!] 
    \caption{This table contains the classification accuracy of three convolutional neural networks when classifying true malware binaries, adversarial examples generated using \cite{Kreuk2018}, and our proposed method. To keep the experiments consistent, we use the \cite{microsoft-kaggle} dataset. We also generate untargeted attacks using AMAO because the method proposed in \cite{Kreuk2018} was designed for a binary detection problem. Attacks on InceptionV3 were black-box attacks while those on LeNet and Simple CNN were white-box.}
    \label{table:defense}
    \centering
    \begin{tabular}{|l||l|l|l|}
    \hline
    \textbf{Model and Data} & None    & Adversarial Training & Distillation \\ \hline
    Simple CNN \& True      & 85.10\% & 84.40\%              & 86.50\%      \\ \hline
    Simple CNN \& \cite{Kreuk18}     & 83.24\% & 83.90\%              & 83.10\%      \\ \hline
    Simple CNN \& AMAO      & 6.50\%  & 7.78\%               & 7.45\%       \\ \hline
    LeNet \& True           & 86.55\% & 86.00\%              & 87.20\%      \\ \hline
    LeNet \& \cite{Kreuk18}          & 85.40\% & 83.75\%              & 85.43\%      \\ \hline
    LeNet \& AMAO           & 6.80\%  & 8.19\%               & 7.30\%       \\ \hline
    InceptionV3 \& True     & 88.45\% & 79.80\%              & 89.10\%      \\ \hline
    InceptionV3 \& \cite{Kreuk18}    & 86.91\%              & 65.0\%  &  66.34\%  \\ \hline
    InceptionV3 \& AMAO     & 26.95\% & 6.20\%               & 6.14\%       \\ \hline
    \end{tabular}
    \end{table}    

    \subsection{Attack analysis}
    The results of both black-box and white-box experiments are given in Table \ref{table:results}. It is interesting to note that one loop of AMAO resulted in malware samples of which only around 50\% were able to successfully fool the InceptionV3 and MalConv. However, the same samples were much more effective against our boosted trees classifier, achieving ~98\% misclassification rate. Our method also consistently outperformed randomized dummy code obfuscation.
    
    It is also interesting to note that the initial highest performing classifier is not necessarily the most robust against adversarial examples. 
    InceptionV3 was marginally better than our simple CNN with respect to classification accuracy, but proved to be much more robust against our executable adversarial examples, maintaining a 50\% and 27\% classification accuracy against single-loop and full AMAO, respectively.
    
    \subsection{Effectiveness against defense strategies}
    To further analyze the effectiveness of our proposed method, we present experimental results of generated adversarial malware examples against popular adversarial defense strategies in Table \ref{table:defense}. We also compare our work with the payload method proposed by \cite{Kreuk2018}.
    
    We present the classification accuracies of a simple convolutional neural network, LeNet, and InceptionV3 when classifying unobfuscated malware, adversarial malware examples generated using the payload method, and adversarial malware examples generated using AMAO. Each model attempted the classification task with no defense, adversarial training, and defensive distillation. The accuracy of classifying true data samples is also included to show that the defense techniques are not overfitting to the adversarial examples. 
    
    The defenses were both done as a "one time" retraining. Each classifier was first trained on true samples before generating adversarial examples to be used in the retraining phase. During retraining, each classifier was trained until the validation accuracy reached the same accuracy as for classifying true samples. We present the classification accuracy on new adversarial examples that were generated against the defended networks. 
    
    Interestingly, defensive distillation slightly increases the classification accuracy of the true test set. The results show that distillation increased the average classification accuracy of the true test set by $1\%$ across the three classifiers.
    
    The defense techniques were not able to significantly increase classification accuracy against attacks generated using AMAO. Surprisingly, InceptionV3's classification drastically dropped when it was trained using adversarial training and distillation. 
    
    The payload method did not reduce any model's classification accuracy significantly. We believe that the addition of bytes to the end of the malware binaries may work with deep neural networks with an embedding layer, such as MalConv, but is not enough to affect image classifiers \cite{Raff17}.  
    
    We also attempted to experimentally compare our method with the method proposed in \cite{Demetrio19}. A toy implementation of the proposed method was tested. However, because this work creates adversarial malware examples by perturbing bytes in the header, we did not include these results as neither the Malimg dataset nor the Microsoft malware challenge dataset contains complete headers. The toy implementation was tested on randomly generated bytes acting as a header to the existing data. Because we had no guarantees that the toy implementation behaved on random byte sequences as the authors intended, we omit these results. However, because of its similarity to the payload method, we believe perturbing bytes in the header only works for architectures similar to MalConv and not image based classifies.

    \subsection{Comparisons to other obfuscation methods}
    In this section, we will begin with a brief overview of some other obfuscation methods (found in \cite{Survey} and the references therein) and compare them to AMAO.
    
    \subsubsection{Subroutine reordering}
    In this type of obfuscation, the order in which the subroutines appear in the executable are permuted. This change does not alter the program logic as it does not affect the execution of a program. 

    \subsubsection{Mixing control flow}
    Control flow mixing relies heavily on the \textit{jmp} instruction to alter the code sequencing but preserve the original behavior. The number of additional \textit{jmp} instruction increases with the number of lines that are permuted, leading to different signatures and different memory mappings. Below is a simple example of control flow mixing.
    \begin{verbatim}
xor eax, eax         00 jmp short 16
mov eax, 0x45        05 mov eax, 0x45
mov ecx, 0x20   -->  0a mov ecx, 0x20
ret                  0f jmp short 1d 
                     16 xor eax, eax
                     1b jmp 05
                     1d ret
    \end{verbatim}
    
    \subsubsection{Instruction substitution}
    Instruction Substitution relies on the fact that instructions can be executed in multiple ways. For example, an \textit{XOR} gate can be constructed in at least 4 ways. There are two common ways to utilize this transformation. 
    
    \textbf{a.} \textit{Static value obfuscation}
    The first is to obfuscate static values that are assigned to variables or moved to registers. We can change a \textit{mov} instruction into a sequence of instructions that is logically equivalent like shown in the example below.
\begin{verbatim}
                    mov eax, 0x45
mov eax, 0x12  -->  add eax, 0x03
                    xor eax, 0x5a
\end{verbatim}
    The sequence on the right results in the same value in eax, but does not immediately leak the value of 0x12.
    
    \textbf{b.} \textit{Binary operators}
    The second commonly used method is to replace standard binary operators, such as addition and subtraction, with more complicated, but equivalent, code sequences. For example, a simple addition $a = b + c$ can be substituted with
\begin{verbatim}
    r = rand()
    a = b + r;
    a = a + c;
    a = a - r;
\end{verbatim}

    \subsubsection{Obfuscation results}
    To test the effectiveness of the obfuscation techniques in a real world setting, we apply obfuscation to the test set given in the Microsoft malware challenge \cite{microsoft-kaggle} and feed it to a trained classifier. In this set of experiments, the top scoring submission to the Microsoft malware challenge was used as the classifier. The combination of features used in this classifier makes it more similar to current malware detection and classification approaches.
    
    \begin{table}[ht!]
    \centering
    \caption{This table compares accuracy scores of the top scoring Microsoft malware challenge \cite{microsoft-kaggle} model trained on the competition malware training dataset and tested on malware data that has undergone different obfuscation.} % expand this part.
    \begin{tabular}{|l ||l|}
    \hline
    \textbf{Obfuscation} & \textbf{Classification Accuracy} \\ \hline \hline
    No Obfuscation & $98.62 \pm 0.21\%$ \\ \hline
    Instruction Sub. & $97.36 \% \pm 0.36\%$ \\ \hline
    Subroutine Reorder & $98.45 \pm 0.61\%$ \\ \hline
    Static Value Obf. & $97.52 \pm 0.29\%$ \\ \hline
    Dummy Code & $87.98 \pm 3.49\%$ \\ \hline
    Mixing Flow & $39.46 \pm 3.70\%$ \\ \hline
    AMAO & $0.0\%$ \\ \hline
    \end{tabular}
    \label{tab:res1}
    \end{table}   

    Table \ref{tab:res1} shows the classification accuracy of the model when fed malware data that has undergone different obfuscation. All obfuscation, other than AMAO, was applied \textbf{randomly} to the malware samples with a fixed file size limit. The results show that instruction substitution, subroutine reordering, and static value obfuscation do not significantly reduce the classification accuracy of the classifier. Random dummy code insertion results in $87.98\%$ classification accuracy, an $11.3\%$ decrease in classification accuracy compared to the same model classifying the original test set. Mixing control flow is most effective out of the randomly inserted obfuscation with a $39.46\%$ classification accuracy. However, as shown in Table \ref{tab:res1} and Section \ref{Experimental results}, AMAO results in $0.0\%$ classification accuracy. AMAO only uses dummy code insertion, but still results in greater evasion rates than the other obfuscation techniques. In Section \ref{Discussion}, we will further discuss the patterns learned by AMAO.

    \subsubsection{A note on packers}
    Packers are the most popular obfuscation technique currently used to evade real-world antivirus systems \cite{Li2010}. Packers use compression algorithms to obfuscate malicious code by attempting to eliminate malign features. Typically, packed malware will contain the compressed malware and a decryption engine that will undo the compression at runtime. We experimented with packed malware but the results were omitted because recent work has shown that entropy, section hashing, and consistently executing graph mining can be used to classify packed malware \cite{Li19, Gibert2018, Shiel2019}. Being able to be detected using entropy and hashing approaches leads us to believe packing is not well-suited for use in adversarial example generation.
    
\section{Discussion}\label{Discussion}
By using obfuscation, our method ensures the adversarial example is executable. This makes our examples effective against dynamic analysis as well as static analysis. A simple method to detect adversarial examples given malware samples to classify would be to run the sample in a sandbox environment to extract dynamic features. If the sample does not run or does not behave maliciously before any kind of termination, it can be categorized as an adversarial example or a corrupted sample. To the best of our knowledge, AMAO is the first proposed method to use obfuscation to create executable adversarial examples. This allows AMAO to modify any section of a binary file. Its resulting malware samples can also be used as input to other detectors and classifiers because both static and dynamic features can be extracted from it. We also believe that the addition of \textit{nop}-like system and API calls, such as \textit{time()}, creates adversarial input to non-image classification models, meaning the adversarial examples cover multiple feature spaces, unlike other methods in the literature. Another example would be to insert \textit{nops}, affecting classifiers that use \textit{n}-grams as a feature.

Obfuscation is usually applied to binaries randomly or without discretion because its goal is only to convolute the program's logic, hiding it from reverse engineers and analysis tools. However, AMAO uses generated adversarial examples, which are created using established methods, to generate obfuscation. Because the obfuscation is designed according to adversarial examples, AMAO is much more effective against classifiers than randomized obfuscation methods.

\subsection{Learning to insert dummy code}
We can further learn effective dummy code insertion from creating executable adversarial malware examples using AMAO by extracting learned patterns.

\begin{figure}[ht]
    \centering
    \begin{subfigure}
        \centering
        \includegraphics[width=0.84\linewidth]{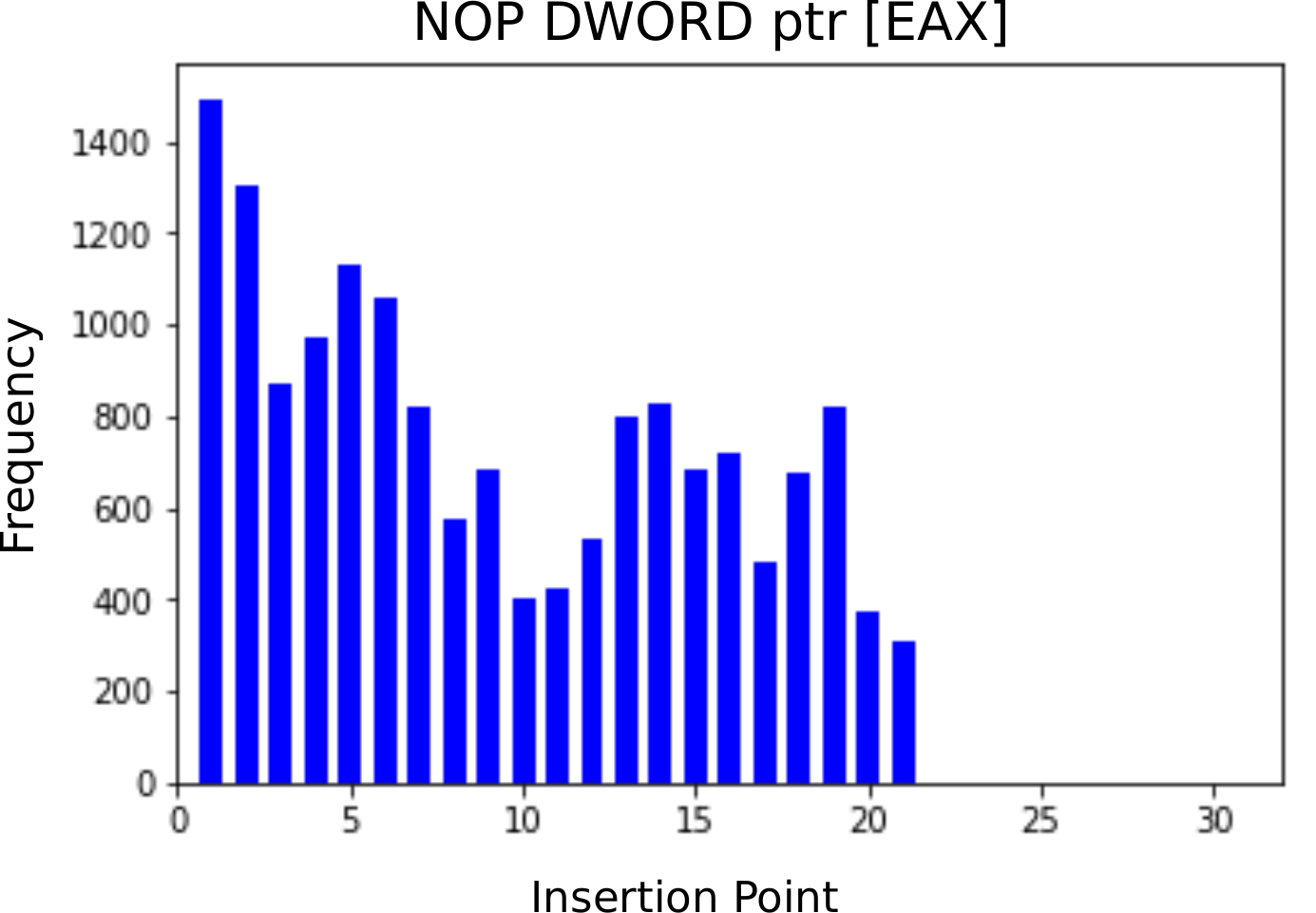}
    \end{subfigure}
    \vspace{1mm}
    \begin{subfigure}
        \centering
        \includegraphics[width=0.815\linewidth]{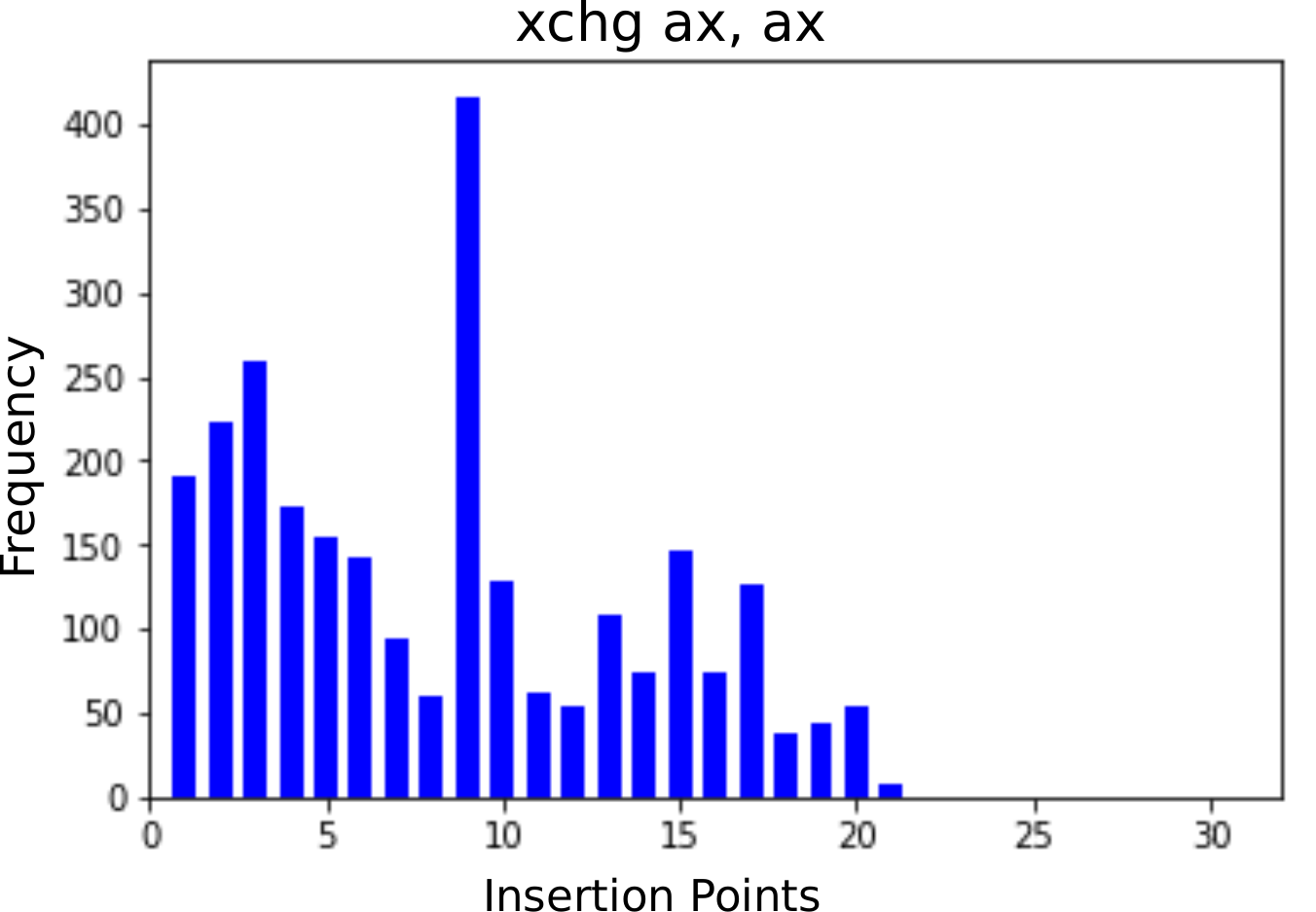}
    \end{subfigure}
    \vspace{1mm}
    \begin{subfigure}
        \centering
        \includegraphics[width=0.79\linewidth]{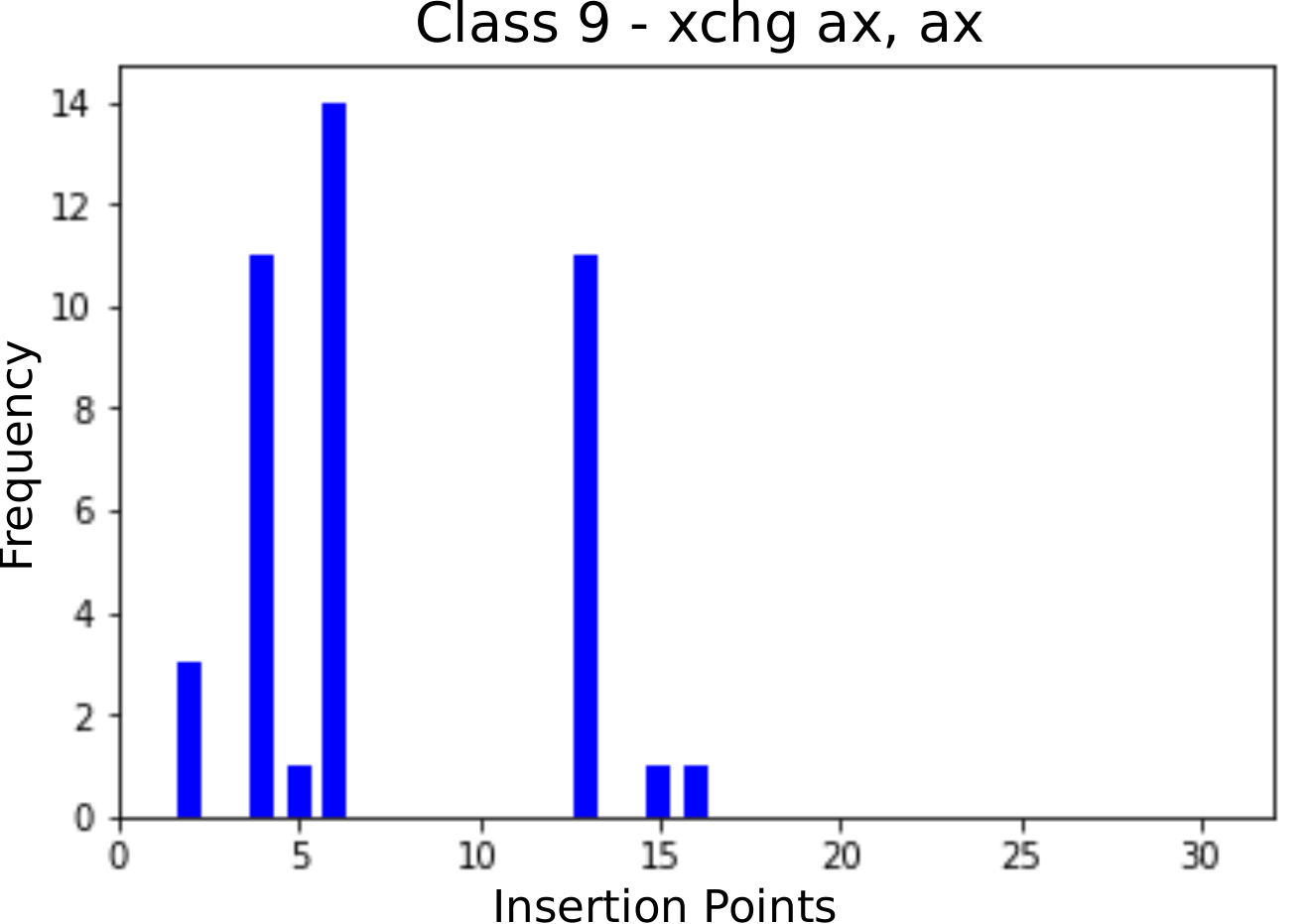}
    \end{subfigure}
    \caption{Obfuscation decision by AMAO was collected to create these three graphs. Each graph shows the frequency a semantic \textit{nop} was inserted at each insertion point. Insertion points are locations in the binary where instructions can be inserted without changing the program logic. Heuristics can be developed using AMAO for general obfuscation or targeting specific classes of malware, e.g. how to insert \textit{xchg ax, ax} to target class 9. }
    \label{fig:learn_ob}
\end{figure}

By outputting the decisions of the dynamic programming algorithm, we can develop a heuristic for dummy code insertion. Specifically, for each semantic \textit{nop} in our vocabulary, we extract the frequency that it is inserted over the set of all valid insertion points. The frequencies of insertion can be separated by criteria such as attack target class, specific semantic-\textit{nops}, and the intersection of the two as can be seen in Figure \ref{fig:learn_ob}.

\section{Related works in adversarial examples}\label{Related works}
Reference \cite{Grosse16} proposed generating adversarial examples of the DREBIN Android malware dataset and claimed up to an 84\% misclassification rate. The DREBIN dataset, introduced in \cite{Drebin}, is a collection of features from static and dynamic analysis of the Android malware. Similar to the work by Grosse, \cite{Hu17} proposed a method to craft adversarial examples against RNNs, further building upon the work of \cite{Papernot15}, which introduced an iterative algorithm to generate adversarial examples and then extended their work to RNNs. Reference \cite{Hu17b} introduced MalGAN, a generative adversarial network used to generate adversarial malware examples to fool black-box machine learning models, and showed high evasion rates. MalGAN currently uses API calls as features, but extending the method to create executable adversarial malware examples would make it a more dangerous tool necessitating increased research towards robust models.

Our approaches are similar in that they are both based around perturbing an input $X$ while limiting insertions of features to indices in a set $I$, which contains locations where insertions will not destroying the application's functionality. However, unlike previous work, our method inserts executable assembly code to the malware, ensuring that binary is immediately executable instead of perturbing extracted features. 

After \cite{Raff17} first proposed detecting malware using sequences of raw bytes with MalConv, \cite{Kreuk19} proposed a method to generate adversarial examples while preserving functionality by injecting a payload of byte sequences into unused sections in the middle or end of a file, and claimed adversarial examples generated using FGSM might not preserve the functionality of the file as the changes to the embedding layer of their network architecture may have unwanted effects on the functionality of the code. Our method circumvents this by using an image representation of malware instead of the one-hot representation used in \cite{Kreuk18}. Our method also is not limited to inserting payloads or adversarial perturbations into non-executable sections as our perturbations are in the form of already executable instructions. This enables our method to also be effective against classifiers using features such as API and system calls, as well as \textit{n}-grams. We also showed in Section \ref{Experimental results} that this payload method cannot evade image-based classifiers such as LeNet and InceptionV3, however AMAO evades classification by MalConv.

Reference \cite{Demetrio19} proposed a method to generate adversarial malware examples by perturbing bytes in the malicious binary's header. The proposed attack uses feature attribution to identify important input features to perturb. This proposed approach is similar to that of \cite{Kreuk18} as they both perturb "allowable" bytes. The payload method perturbs bytes in the data sections or to padding at the end of the binary and \cite{Demetrio19}'s method perturbs bytes in the header that are unimportant to the header's functionality. Because their similarities, header perturbations suffer the same drawbacks as the payload method.

%%%%%%%%%%%%%%%%%%%%%%%%%%%%%%%%%%%%%%%%%%%%%%
%% TODO: add we beat malconv, they don't beat
%%       img classifiers
%%%%%%%%%%%%%%%%%%%%%%%%%%%%%%%%%%%%%%%%%%%%%%

\section{Conclusion and future work}\label{conclusion}
In this paper, we introduced a novel method for generating executable adversarial examples of malware samples. We designed and used the Adversarial Malware Alignment Obfuscation algorithm and showed, through our evaluation on the Malimg dataset, that we were able to achieve high evasion rates in both white and black box settings against state of the art image and malware classifiers. Our method works by creating adversarial examples and then designing obfuscation based off of those examples. To the best of our knowledge, AMAO is the first proposed method to incorporate obfuscation in creating executable adversarial malware examples. This allows adversarial perturbation in any section of the malware while ensuring its executability and persevering its functionality. Because AMAO works with byte sequences, the resulting examples can be easily converted to binary files from images. This means that our adversarial examples can also be used against classifiers that use static and dynamic features extracted using program analysis techniques. 

We further showed the effectiveness of our proposed method by evading deep neural network classifiers such as LeNet5 and InceptionV3 with and without popular defense techniques, adversarial training and distillation. In this defensive analysis, we compared the proposed method with that of \cite{Kreuk2018} and showed that our proposed method achieved higher evasion rates against image-based classifiers while also evading other classifiers such as MalConv. 

This work also introduced an effective obfuscation algorithm to evade classification to replace ad-hoc or randomized obfuscation. By gathering information during the proposed adversarial example generation algorithm, heuristics for obfuscating malware binaries can be developed. Executable adversarial malware examples can be generated by using the learned patterns as a guide for semantic \textit{nop} insertion. 

\bibliographystyle{IEEEtran}
\bibliography{ms}

\end{document}